\documentclass[11pt]{article}
\pdfoutput=1
\usepackage{jheppub}
\usepackage{tabu}
\usepackage[vcentermath]{youngtab}
\usepackage[usenames,dvipsnames,table]{xcolor}
\usepackage{graphicx,amsmath,amssymb,amsthm,multirow,array,bm,bbm,esint}
\usepackage[mathscr]{eucal}
\usepackage[bbgreekl]{mathbbol}
\usepackage{epsf,amsfonts}
\usepackage{slashed}
\usepackage[numbers,sort&compress]{natbib}
\usepackage{minibox}
\usepackage{rotating}
\usepackage{pdflscape}
\usepackage{array,tikz-cd}
\usepackage{framed}
\usepackage{mathtools}
\usepackage{tikz} 
\usepackage[compat=1.1.0]{tikz-feynman}
\usetikzlibrary{snakes}
\usepackage{float}
\usepackage{verbatim}
\usepackage{subfig}
\usepackage[export]{adjustbox}

\usepackage{slashed}

\newcommand{\diff}{\text{Diff}(S^1) }
\newcommand{\cF}{\mathcal{F}}
\newcommand{\dr}{\text{\tiny dressed}}

\def\res{\mathop{\text{Res}}\limits}        

\newcommand{\ie}{\emph{i.e.}, }
\newcommand{\eg}{\emph{e.g.}, }

\makeatletter
\newcommand{\doubletilde}[1]{{%
  \mathpalette\double@tilde{#1}%
}}
\newcommand{\double@tilde}[2]{%
  \sbox\z@{$\m@th#1\tilde{#2}$}%
  \ht\z@=.9\ht\z@
  \tilde{\box\z@}%
}
\makeatother

\title{Quantum Correction to Chaos in Schwarzian Theory}

\author[a]{Yong-Hui Qi}
\author[b]{\!, Sang-Jin Sin}
\author[c]{\!, Junggi Yoon}
\affiliation[\,a,b]{Department of Physics, Hanyang University, Seoul, 04763, Korea}
\affiliation[\,c]{School of Physics, Korea Institute for Advanced Study\\
85 Hoegiro Dongdaemun-gu, Seoul 02455, Republic of Korea.}
\emailAdd{yhqi@pku.edu.cn}
\emailAdd{sjsin@hanyang.ac.kr}
\emailAdd{junggiyoon@kias.re.kr}

\preprint{{\raggedleft \tt KIAS-P19031 \par} }

\abstract{We discuss the quantum correction to chaos in the Schwarzian theory. We carry out the semi-classical analysis of the Schwarzian theory to study Feynman diagrams of the Schwarzian soft mode. We evaluate the contribution of the soft mode to the out-of-time-order correlator up to order $\mathcal{O}(g^4)$. We show that the quantum correction of order $\mathcal{O}(g^4)$ by the soft mode decreases the maximum Lyapunov exponent ${2\pi \over \beta}$. }

\begin{document}
\maketitle

\section{Introduction}
\label{sec:introduction}

Recently, quantum chaos has been intensively investigated in AdS/CFT. One of the diagnosis of the chaos is the well-known butterfly effect which stands for the sensitivity of a system on the initial condition. In quantum system, such sensitivity can be captured by the out-of-time-ordered correlator~(OTOC) defined by~\cite{Shenker:2013pqa,Shenker:2013yza,Roberts:2014isa,Shenker:2014cwa,Maldacena:2015waa}
\begin{equation}
\langle V(t)W(0)V(t) W(0)\rangle_\beta \ ,
\end{equation}
where $\langle \cdots \rangle_\beta $ denotes the thermal expectation value at temperature $\beta^{-1}$. The sensitivity of a chaotic system on the initial condition leads to an exponential growth of the OTOC between the dissipation time $t_d\sim \beta$ and the scrambling time $t_\ast$~\cite{Shenker:2013pqa,Shenker:2013yza,Roberts:2014isa,Shenker:2014cwa,Maldacena:2015waa}:
\begin{equation}
	{\langle V(t)W(0)V(t) W(0)\rangle_\beta\over \langle V(t)V(t) \rangle_\beta \langle W(0)W(0) \rangle_\beta } = 1- \kappa g^2 e^{\lambda_L t}\hspace{10mm} \mbox{for } \quad t_d<t<t_\ast \ .\label{eq: otoc long time behavior}
\end{equation}
Here, $g^2$ is proportional to the inverse of the large central charge $c$, and $\kappa$ is a constant depending on the details of models. Note that the scrambling time $t_\ast$ is of order ${1\over \lambda_L}\log {1\over g^2}$.  The exponential growth rate $\lambda_L$ is known as the Lyapunov exponent. It was shown~\cite{Maldacena:2015waa} that the Lyapunov exponent $\lambda_L$ is bounded in a quantum field theory with unitarity and causality.\footnote{Also, see~\cite{Yoon:2019cql} for understanding of the bound on chaos from the stability of the Schwarzian theory.} \ie
\begin{equation}
	\lambda_L\; \leqq \; {2\pi \over \beta}\ .
\end{equation}
This bound on chaos indicates the concept of the maximal chaos, and it was shown that SYK models~\cite{Sachdev:1992fk,kitaevfirsttalk,KitaevTalks,Polchinski:2016xgd,Jevicki:2016bwu,Maldacena:2016hyu,Jevicki:2016ito,Gross:2016kjj,Fu:2016vas,Yoon:2017nig,Yoon:2017gut,Narayan:2017hvh,Ferrari:2019ogc}, the tensor models~\cite{Gurau:2010ba,Carrozza:2015adg,Witten:2016iux,Gurau:2016lzk,Klebanov:2016xxf,Yoon:2017nig}, the dilaton gravity on nearly-AdS$_2$~\cite{Maldacena:2016upp} and string worldsheet theories~\cite{deBoer:2017xdk,Murata:2017rbp,Banerjee:2018twd} have the maximal Lyapunov exponent $\lambda_L={2\pi \over \beta}$. In such a maximally chaotic system, \eqref{eq: otoc long time behavior} is almost constant in early time because the exponential growth is negligible compared to the leading constant term. As time is increased, the sub-leading exponential growth with $\lambda_L={2\pi \over \beta}$ in \eqref{eq: otoc long time behavior} is comparable to the leading constant around the scrambling time $t_\ast\sim {1\over \lambda_L}\log {1\over g^2}$. At the same time, the next sub-leading term of order $\mathcal{O}(g^4)$ would also become of importance. Then, it is interesting to ask a question whether the next sub-leading correction of order $\mathcal{O}(g^4)$ increases or decreases the maximum Lyapunov exponent $\lambda_L={2\pi \over \beta}$.

In this paper, we will make an attempt to answer this question\footnote{Note that there has been a series of works to pursue a similar question~\cite{Fitzpatrick:2016thx,Chen:2016cms,Fitzpatrick:2016mjq}. Also, see~\cite{Hijano:2015rla,Perlmutter:2015iya,Hikida:2017byl}.} in the Schwarzian theory~\cite{Stanford:2017thb,Mertens:2017mtv,Mertens:2018fds,Qi:2019dxj,Lam:2018pvp,Blommaert:2018oro,Cotler:2018zff} which describes the low-energy sector of the SYK-like models and the dilaton gravity on the nearly-AdS$_2$. And, it is responsible for the saturation of the bound on chaos thereof. Although not all the quantum correction of order $\mathcal{O}(g^4)\sim \mathcal{O}({1\over c^2})$ to the OTOC exhibits universality, the contribution of the Schwarzian modes of order $\mathcal{O}(g^4)$ will be universal.

The outline of this paper is as follows. {\bf In Section~\ref{sec:  action}}, we review the semi-classical analysis of the Schwarzian theory. Then, we evaluate the propagator of the Schwarzian soft mode and its loop correction. {\bf In Section~\ref{sec:  otoc}}, we consider a bi-local field and its Euclidean two point function which corresponds to four point function of a fundamental local field. By studying the soft mode expansion of the dressed bi-locals, we evaluate the contribution of the soft mode to the Euclidean two point function of bi-locals. Then, taking analytic continuation to the real time, we obtain the soft mode contribution to OTOC. {\bf In Section~\ref{sec:conclusion}}, we make concluding remarks, and we present caveats and future directions.

\section{Schwarzian Theory}
\label{sec:  action}

\subsection{Review}
\label{sec:  review}

We begin with the review of the Schwarzian theory in~\cite{Stanford:2017thb}. The partition function of the Schwarzian theory is given by
\begin{equation}
	Z[g]=\int {\mu[\phi]\over SL(2,\mathbb{R}) }\;  \exp \left[ -{1\over 2g^2} \int_0^{2\pi} d\tau \left( \left({\phi''\over \phi'}\right)^2- (\phi')^2 \right) \right]\ ,\label{def: partition function of schwarzian with measure}
\end{equation}
where $\phi(\tau)\in \diff$ is a diffeomorphism of a circle. The Schwarzian theory has $SL(2,\mathbb{R})$ symmetry  given by
\begin{equation}
	\tan {\phi\over 2}\quad\longrightarrow \quad {a \tan{\phi\over 2} + b\over c \tan{\phi\over 2}+d}\hspace{10mm}(a,b,c,d\in \mathbb{R}\quad\text{and}\quad ad-bc=1)\ ,
\end{equation}
and we mod out the $SL(2,\mathbb{R})$ volume. Hence, the physical degree of freedom of the Schwarzian theory lives on the quotient space $\text{Diff}(S^1)/SL(2)$. Note that $\mu[\phi]$ is the reparametrization invariant measure. After $SL(2,\mathbb{R})$ gauge-fixing, the measure becomes~\cite{Bagrets:2016cdf,Stanford:2017thb,Saad:2019lba}
\begin{equation}
	\mu[\phi] \equiv{\mathcal{D}\phi\over  { \displaystyle\prod_\tau} \phi'(\tau)}\ .
\end{equation}
The measure $\mu[\phi]$ can be exponentiated by introducing a fermion $\psi(\tau)$, and the partition function of the Schwarzian in~\eqref{def: partition function of schwarzian with measure} can be written as~\cite{Stanford:2017thb}
\begin{align}
	Z[g]=\int { \mathcal{D}\phi\mathcal{D}\psi\over SL(2,\mathbb{R})} \; e^{-S}\ ,
\end{align}
where the action $S$ is given by
\begin{align}
	S={1\over 2} \int_0^{2\pi} d\tau \left[ {1\over g^2}\left({\phi''\over \phi'}\right)^2-{1\over g^2} (\phi')^2 +{\psi'' \psi'\over (\phi')^2}-\psi' \psi \right]\ .\label{eq: schwarzian action}
\end{align}

In the weak coupling limit $g\ll 1$, one can perform the semi-classical analysis of the action in~\eqref{eq: schwarzian action} by expanding the diffeomorphism $\phi(\tau)$ around a saddle point $\phi(\tau)=\tau$:
\begin{equation}
	\phi(\tau)= \tau+ g\; \epsilon(\tau)\ ,
\end{equation}
Accordingly, the Schwarzian action in \eqref{eq: schwarzian action} can be expanded with respect to $g$:
\begin{align}
	S=-{\pi\over g^2} + S^{(2)} + g S^{(3)}+g^2 S^{(4)}+\mathcal{O}(g^3)\ ,
\end{align}
where we have
\begin{align}
	S^{(2)}=&{1\over 2} \int_0^{2\pi }d\tau \;\left[ (\epsilon'')^2 -(\epsilon')^2 + \psi''\psi' -\psi'\psi \right]\ ,\label{eq: quadratic action}\\
	S^{(3)}=&{1\over 2} \int_0^{2\pi }d\tau \;\epsilon'\left[-2(\epsilon'')^2 - 2\psi''\psi'  \right]\ ,\label{eq: cubic action}\\
	S^{(4)}=&{1\over 2} \int_0^{2\pi }d\tau \;(\epsilon')^2\left[ 3(\epsilon'')^2 + 3\psi''\psi'  \right]\ .\label{eq: quartic action}
\end{align}
After fixing the $SL(2)$ gauge~\cite{Stanford:2017thb}
\begin{equation}
\int d\tau\; \epsilon(\tau)=\int d\tau \;e^{\pm i \tau } \epsilon(\tau)=0\qquad,\qquad \int d\tau \; \psi(\tau)=\int d\tau \; e^{\pm i \tau} \psi(\tau)=0\ ,
\end{equation}
we Fourier-transform the soft mode $\epsilon(\tau)$ and the fermion $\psi(\tau)$ to the (discrete) momentum space:
\begin{equation}
	\epsilon(\tau)= \sum_{|n|\geqq 2} \epsilon_n e^{-i n \tau}\hspace{6mm},\hspace{6mm} \psi(\tau)= \sum_{|n|\geqq 2} \psi_n e^{-i n \tau}\ .
\end{equation}
In the momentum space, \eqref{eq: quadratic action}$\sim$\eqref{eq: quartic action} can be written as
\begin{align}
	S^{(2)}=&\pi \sum_{|n|\geqq 2}  n^2(n^2-1)\epsilon_{-n}\epsilon_n - \pi i \sum_{|n|\geqq 2}  n(n^2-1)\psi_{-n}\psi_n\ , \label{eq: quad action momentum}\\
	S^{(3)}=&{2\pi i\over 3} \sum_{\substack{|n|,|m|\geqq 2\\ |m+n|\geqq2}}  mn(m+n)(m^2+mn+n^2) \epsilon_{-m-n}\epsilon_m\epsilon_n \cr
	&+  \pi \sum_{\substack{|n|,|m|\geqq 2\\ |m+n|\geqq2}}  nm(m+n)(n+2m)\psi_{-m-n}\psi_m \epsilon_n\ ,  \label{eq: cubi action momentum} \\
	S^{(4)}=& -{\pi\over 2}  \sum_{\substack{|n|,|m|,|p|\geqq 2\\ |m+n+p|\geqq2}}  mnp(m+n+p)(m^2+n^2+p^2+mn+np+pm) \epsilon_{-m-n-p}\epsilon_m\epsilon_n\epsilon_p \cr
	&+ {3\pi i\over 2}  \sum_{\substack{|n|,|m|,|p|\geqq 2\\ |m+n+p|\geqq2}}  mn p (m+n+p)(m+n+2p)\psi_{-m-n-p}\psi_p \epsilon_m \epsilon_n \ .\label{eq: quartic action momentum}
\end{align}
\begin{figure}[t!]
\centering
\begin{subfloat}[][Soft mode]{
\centering
$\langle \epsilon_{-n} \epsilon_n \rangle_{\text{\tiny free}}\;\;= $
 \includegraphics[width=.2\linewidth,valign=c]{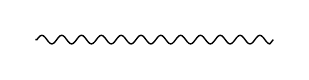}}
\end{subfloat}
\quad
\begin{subfloat}[][Fermion]{
\centering
$\langle \psi_{-n} \psi_n \rangle_{\text{\tiny free}}\;\;= $
\includegraphics[width=.2\linewidth,valign=c]{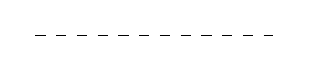}}
\end{subfloat}

	\caption{Free propagator of the Schwarzian soft mode and the fermion.}
	\label{fig: propagator}
\end{figure} 
\begin{figure}[t!]
\centering
\begin{subfloat}[][Cubic vertex]{\includegraphics[width=.23\linewidth,valign=b]{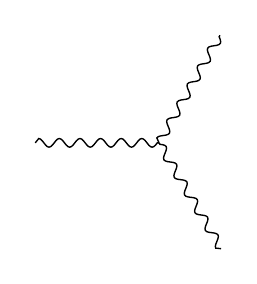}}
\end{subfloat}
\begin{subfloat}[][Fermi cubic vertex]{\includegraphics[width=.23\linewidth,valign=b]{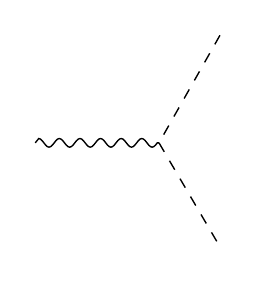}}
\end{subfloat}
\quad
\begin{subfloat}[][Quartic vertex]{\includegraphics[width=.23\linewidth,valign=b]{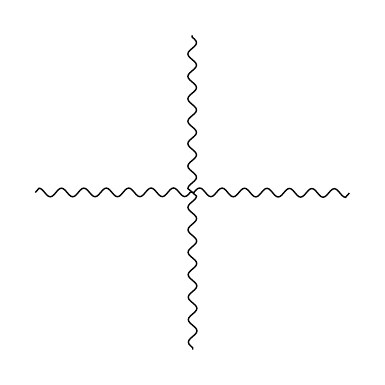}}
\end{subfloat}
\begin{subfloat}[][Fermi quartic vertex]{\includegraphics[width=.23\linewidth,valign=b]{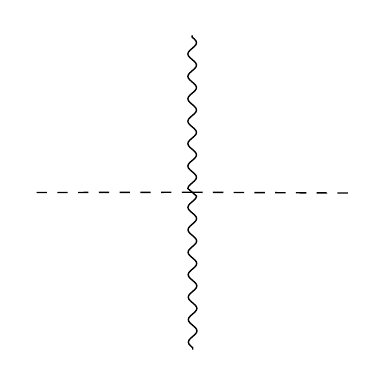}}
\end{subfloat}
	\caption{Vertices}
	\label{fig: vertices}
\end{figure} 
They give the free propagator as well as cubic, quartic vertices of the soft modes and fermions (see Figure~\ref{fig: propagator}$\;\sim\;$\ref{fig: vertices}). For the leading quantum correction to OTOC, it is enough to consider the interactions up to quartic vertex.

\subsection{Soft Mode Propagator}
\label{sec:  soft mode propagator}

From the quadratic action in~\eqref{eq: quad action momentum}, one can read off the free propagator of the soft mode and the fermion in (discrete) momentum space:
\begin{align}
	\langle \epsilon_{-n} \epsilon_n \rangle_{\text{\tiny free}} = {1\over 2\pi } {1\over n^2(n^2-1)}\hspace{5mm},\hspace{8mm} \langle \psi_{-n} \psi_n \rangle_{\text{\tiny free}} = {1\over 2\pi i} {1\over n(n^2-1)}\ .\label{eq: leading soft mode propagator}
\end{align}
Now, we will evaluate the loop correction of order $\mathcal{O}(g^2)$ to the propagators in \eqref{eq: leading soft mode propagator}. At order $\mathcal{O}(g^2)$, there are two types of loops: one with a quartic vertex and one with two cubic vertices. And, each loop correction gives divergence. However, as in the calculation of the free energy~\cite{Stanford:2017thb}, the divergence in the bosonic loop is cancelled with that of the fermion loop of the same type, which leads to a finite answer.

\begin{figure}[t!]
\centering
\begin{subfloat}[][Soft mode loop]{
$\langle \epsilon_{-n} \epsilon_n \rangle_{\text{\tiny quartic}}\;\;= $
 \includegraphics[width=.25\linewidth,valign=c]{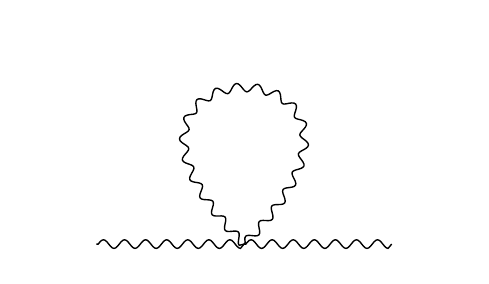}}
\end{subfloat}
\qquad
\begin{subfloat}[][Fermi loop]{$\langle \psi_{-n} \psi_n \rangle_{\text{\tiny quartic}}\;\;= $
\includegraphics[width=.25\linewidth,valign=c]{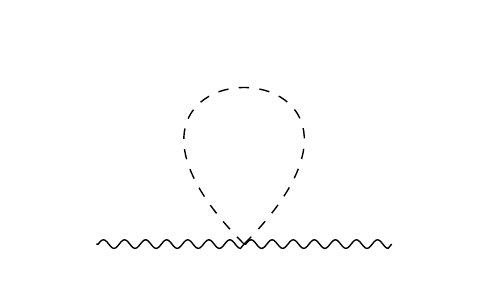}}
\end{subfloat}

	\caption{The loop correction by one quartic vertex to the propagator.}
	\label{fig: quartic loop correction}
\end{figure} 

\vspace{3mm}
\noindent
{\bf One-loop with One Quartic Vertex:} We evaluate the contribution of the two diagrams in Figure~\ref{fig: quartic loop correction} with bosonic and fermi loops made of quartic vertex:
\begin{equation}
	\langle \epsilon_{-n} \epsilon_n \rangle_{\text{\tiny quartic}} 
	= -{6g^2\over 8\pi^2} \sum_{|m|\geqq 2}{ n^2+m^2\over n^2(n^2-1)^2(m^2-1)  } +{6g^2\over 8\pi^2} \sum_{|m|\geqq 2}{m^2\over n^2(n^2-1)^2(m^2-1)}\ .
\end{equation}
Note that each series does not converge because the term of each series does not converge to zero as $m$ goes to infinity. However, the summation of two terms converges to zero as $m\rightarrow \infty$, and its series can be expressed as contour integral as follow. 
\begin{align}
	\langle \epsilon_{-n} \epsilon_n \rangle_{\text{\tiny quartic}} 
	=&-{3g^2\over 4\pi^2 } \sum_{|m|\geqq 2}{1\over (m^2-1)(n^2-1)^2} \ ,\\
	=& -{3g^2\over4\pi^2}{1\over (n^2-1)^2} {1\over 2\pi i}\oint_{\mathcal{C} } d\zeta\;  {\pi \over \tan \pi \zeta } {1\over (\zeta^2-1)}\ .
\end{align}
where the contour $\mathcal{C}$ is a collection of small counterclockwise circles centered at $\zeta \in \mathbb{Z}/\{-1,0,1\}$. By deforming the contour, it can be changed into the sum of the residues at $\zeta=-1,0,1$:
\begin{equation}
	\langle \epsilon_{-n} \epsilon_n \rangle_{\text{\tiny quartic}} = {3g^2\over 4\pi^2}{1\over (n^2-1)^2}  \sum_{m=-1,0,1} \res_{\zeta = m} {\pi \over \tan \pi \zeta } {1\over (\zeta^2-1)}=- {9g^2\over 8\pi^2}{1\over (n^2-1)^2} \ . \label{eq: quartic loop correction}
\end{equation}

\begin{figure}[t!]
\centering
\begin{subfloat}[][Soft mode loop]{$\langle \epsilon_{-n} \epsilon_n \rangle_{\text{\tiny cubic}}\;\;= $
 \includegraphics[width=.25\linewidth,valign=c]{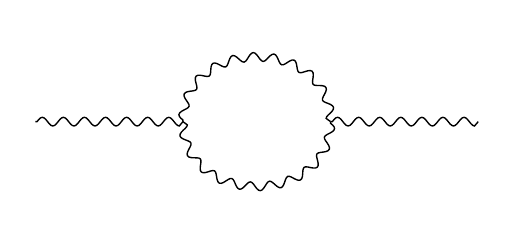}}
\end{subfloat}
\qquad
\begin{subfloat}[][Fermi loop]{$\langle \psi_{-n} \psi_n \rangle_{\text{\tiny cubic}}\;\;= $
\includegraphics[width=.25\linewidth,valign=c]{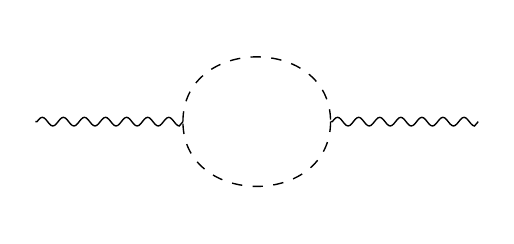}}
\end{subfloat}
	\caption{The loop correction by two cubic vertices to the propagator.}
	\label{fig: loop correction cubic vertex}
\end{figure} 

\noindent
{\bf One-loop with Two Cubic Vertices:} In a similar way, one can calculate the sum of two diagrams with bosonic and fermi loop composed of two cubic vertices in Figure~\ref{fig: loop correction cubic vertex}. Again, each diagram does not converge, but the summation of the two diagram gives a finite contribution.
\begin{align}
	\langle \epsilon_{-n} \epsilon_n \rangle_{\text{\tiny cubic}} 
	=&{g^2\over 2\pi^2} \sum_{\substack{|m|\geqq 2 \\ |m+n|\geqq 2}} {(m^2+m n +n^2)^2\over n^2(n^2-1)^2(m^2-1)((m+n)^2-1)}\cr
	&-{g^2\over 8\pi^2} \sum_{\substack{|m|\geqq 2 \\ |m+n|\geqq 2}} {m(n+m)(n+2m)^2\over n^2(n^2-1)^2 (m^2-1)((m+n)^2-1)}\ ,\cr
	=&{g^2\over 8\pi^2} \sum_{\substack{|m|\geqq 2 \\ |m+n|\geqq 2}} {7m^2+7mn+4n^2\over (n^2-1)^2(m^2-1)((m+n)^2-1)}\ .
\end{align}
It can also be written as contour integral as before. 
\begin{align}
	\langle \epsilon_{-n} \epsilon_n \rangle_{\text{\tiny cubic}} ={g^2\over 8\pi^2} \sum_{\substack{|m|\geqq 2 \\ |m+n|\geqq 2}} {1\over (n^2-1)^2} {1\over 2\pi i } \oint_{\mathcal{C}} {\pi \over \tan \pi \zeta}  {7\zeta^2+7 \zeta n+4n^2\over (\zeta^2-1)((\zeta+n)^2-1)}\ .
\end{align}
However, one has to consider the contour for $|n|> 2$ case and for $n=2$ case separately. For the case of $|n|> 2$, the contour $\mathcal{C}$ is a collection of small counterclockwise circles centered at $\zeta \in \mathbb{Z}/\{-n-1,-n,-n+1,-1,0,1\}$ . For $n=2$, the contour $\mathcal{C}$ is a collection of small counterclockwise circles around integers except for $\{-3,-2,-1,0,1\}$ \ie $\zeta \in \mathbb{Z}/\{-3,-2,-1,0,1\}$, and similar for $n=-2$. (See Figure~\ref{fig: contour2}) By deforming the contour, we pick up the residue at the rest of the poles, and we have
\begin{align}
	\langle \epsilon_{-n} \epsilon_n \rangle_{\text{\tiny cubic}} 
	=\begin{cases}
	 {g^2\over 8\pi^2 } {-112+104n^2+179n^4-111n^6 + 12n^8\over n^2(n^2-4)^2(n^2-1)^3} \hspace{3mm}&\text{for}\quad n\ne \pm 2 \\
	-{g^2\over 96} +{1465 g^2\over 6912 \pi^2}  &\text{for}\quad n=\pm2\\
	\end{cases}\ .\label{eq: cubic loop correction}
\end{align}

\newpage

\section{Out-of-time-ordered Correlator}
\label{sec:  otoc}

We will evaluate the contribution of the Schwarzian soft mode to OTOC. For this, we first evaluate Euclidean four point function in a specific configuration, and then we will take analytic continuation to real time OTOC.

\subsection{Dressed Bi-local Field}
\label{sec: dressed bilocal field}

The OTOC is basically a four point function of ``matter'' fields. In the SYK model, the OTOC of the fundamental fermion $\chi^i(\tau)$ $(i=1,2,\cdots, N)$ was evaluated~\cite{Maldacena:2016hyu,Peng:2017spg,Yoon:2017nig,Yoon:2017gut,Narayan:2017hvh}. This can also be viewed as two point function of the bi-local field $\psi(\tau_1,\tau_2)\equiv {1\over N} \sum_{i=1}^N \chi^i(\tau_1)\chi^i(\tau_2)$. A similar bi-local field from a matter scalar field was used to evaluated OTOC in the two-dimensional dilaton gravity on the nearly-AdS$_2$~\cite{Maldacena:2016upp}. Here, the gravity sector is described by Schwarzian action, and the scalar matter field was included on top of the Schwarzian mode. On the other hand, without coupling to an extra matter field, one can also construct the bi-local field by a (boundary-to-boundary) $SL(2)$ Wilson line of BF theory for AdS$_2$~\cite{Blommaert:2018oro,Lam:2018pvp,Blommaert:2018iqz,Mertens:2019tcm,Iliesiu:2019xuh} or Chern- Simons gravity for AdS$_3$~\cite{Jahnke:2019gxr,Narayan:2019ove}.\footnote{For the path integral representation of Wilson line, one need to consider a probe particle moving on the $SL(2,\mathbb{R})$ group manifold~\cite{Ammon:2013hba,Castro:2018srf}.} One can consider a smooth fluctuation around the constant background (\eg BTZ black hole) with a fixed holonomy along the time circle. This fluctuation can be described by Schwarzian action on the boundary~\cite{Cotler:2018zff,Poojary:2018esz,Jahnke:2019gxr,Narayan:2019ove}. For OTOC, one can study Wilson line evaluated with the $SL(2)$ gauge field corresponding to the smoothly fluctuated background, which can be interpreted as bi-local field dressed by the soft mode~\cite{Jahnke:2019gxr,Narayan:2019ove}. Note that this soft mode generates conformal transformation on the boundary. Hence, the dressed bi-local field can equivalently be obtained by the conformal transformation of the (boundary-to-boundary) two point function. Then, one can expand the dressed bi-local field with respect to the soft mode:
\begin{align}
	&{\Phi^{\text{\tiny dressed}}(\tau_1,\tau_2)\over \Phi_{cl}(\tau_1,\tau_2)}\equiv { [\phi'(\tau_1) \phi'(\tau_2)]^h\over \left[\sin {1 \over 2}(\phi(\tau_1) -\phi(\tau_2) )\right]^{2h}}\Bigg/ {1\over [\sin{\tau_{12}\over 2 }]^{2h} }\cr
	=& 1+ g\sum_{|n|\geqq 2} \epsilon_n e^{-in \chi} f^{(1)}_n(\sigma) +g^2  \sum_{|m|,|n|\geqq 2}\epsilon_m \epsilon_n e^{-i (m+n)\chi}f^{(2)}_{m,n}(\sigma) \cr
	&\hspace{10mm}+g^3\sum_{|m|,|n|,|p|\geqq 2} \epsilon_m \epsilon_n\epsilon_p e^{-i(m+n+p)\chi} f^{(3)}_{m,n,p}(\sigma)+\cdots\ ,\label{eq: soft mode expansion}
\end{align}
where $h$ is the conformal dimension of the matter field, and $\Phi_{cl}(\tau_1,\tau_2)$ is the leading term in the soft mode expansion which corresponds to the two point function in the constant background. Here, we defined the center of time $\chi$ and the relative time $\sigma$ to be
\begin{align}
	\chi={1\over 2} (\tau_1+\tau_2) \hspace{5mm},\hspace{8mm} \sigma={1\over 2} (\tau_1-\tau_2)\ .
\end{align}
In addition, the soft mode eigenfunction $f^{(1)}_n(\sigma)$ and $f^{(2)}_{m,n}(\sigma)$ in~\eqref{eq: soft mode expansion} are found to be
\begin{align}
	f^{(1)}_n(\sigma)\equiv& - 2 i  h   \left[n\cos n\sigma -{\sin n\sigma \over \tan \sigma } \right] \ , \\
	f^{(2)}_{m,n}(\sigma)\equiv &-2h^2 \left( m\cos m\sigma-{\sin m\sigma \over \tan \sigma} \right)\left( n\cos n\sigma-{\sin n\sigma \over \tan \sigma} \right)\cr
	&+ h \left[ mn\cos (m+n)\sigma  - {\sin m\sigma \sin n\sigma \over \sin^2 \sigma} \right]\ .
\end{align}
In particular, it is convenient to evaluate them at $\sigma=-{\pi \over 2}$:
\begin{align}
	f^{(1)}_n(-{\pi \over 2})=&-2 i h n \cos {n\pi\over 2} \ ,\label{eq: soft mode1 at special sigma}\\
	f^{(2)}_{m,n}(-{\pi \over 2})=&-h \left[ (2h-1) mn \cos{m\pi \over 2}\cos {n\pi \over 2}+(mn+1) \sin{ m\pi \over 2} \sin{n\pi \over 2} \right]\ . \label{eq: soft mode2 at special sigma}
\end{align}
The form of $f^{(3)}_{m,n,p}(\sigma)$ is complicated, but it is enough to evaluate it at a particular value for our purpose.
\begin{equation}
	f^{(3)}_{m,n,-m}(-{\pi \over 2})=-{1\over 3} i h e^{-in\chi}n \cos{n\pi \over 2}\left[h+2(h-1)^2 m^2 +h( -1 +2(h-1)m^2 )\cos m\pi \right]\ .
\end{equation}
%

\subsection{Euclidean Four Point Function}
\label{sec: four point function}

In this section, we will evaluate the contribution of the soft mode to the Euclidean two point function of the dressed bi-local fields $\Phi^{\text{\tiny dressed}}$, which corresponds to four point function of a matter field:
\begin{equation}
	\langle \Phi^{\text{\tiny dressed}}(\tau_1,\tau_2)\Phi^{\text{\tiny dressed}}(\tau_3,\tau_4) \rangle\ .
\end{equation}
The leading contribution is the product of the one point function of the bi-locals which corresponds to the disconnected diagrams in the four point function of the matters in the constant background.
\begin{equation}
	 \langle \Phi^{\text{\tiny dressed}}(\tau_1,\tau_2)\Phi^{\text{\tiny dressed}}(\tau_3,\tau_4) \rangle=\Phi_{cl}(\tau_1,\tau_2)\Phi_{cl}(\tau_3,\tau_4) +\cdots \ ,
\end{equation}
where the leading one-point function $\Phi_{cl}(\tau_1,\tau_2)$ is given by
\begin{equation}
	\Phi_{cl}(\tau_1,\tau_2)\equiv\langle \Phi^{\text{\tiny dressed}}(\tau_1,\tau_2)\rangle_{\text{\tiny leading}}\sim {1\over \left[\sin { \tau_{12} \over 2}\right]^{2h} } \ . \label{eq: classical one point function}
\end{equation}
Note that one can evaluate the correction to the one point function of the bi-local field. For this, we perform the soft mode expansion of the one point function:
\begin{align}
	\Phi^\dr_0(\sigma)\equiv&{\langle \Phi^\dr(\tau_1,\tau_2) \rangle\over \Phi_{cl}(\tau_1,\tau_2) }= 1 + g^2\sum_{|n|\geqq 2} \langle \epsilon_{-n}\epsilon_n \rangle f^{(2)}_{-n,n}(\sigma)+\cdots\cr
	=&1+ {1\over 4}g h(4+8h-\pi^2)+ \cdots \ . \label{eq: one point function}
\end{align}
Note that $\Phi^\dr_0$ is independent of $\chi$ because of $SL(2)$ charge. In this paper, we will consider a fixed value of $\sigma$ (\ie $\sigma=-{\pi \over 2}$), and then $\Phi^\dr_0$ is nothing but a numerical constant.

One may evaluate the Euclidean four point function for any choice of $(\tau_1,\tau_2;\tau_3,\tau_4)$. However, for the analytic continuation to a particular OTOC, it is enough to consider the following configuration which simplifies the calculation of the four point function:
\begin{align}
	(\tau_1,\tau_2,\tau_3,\tau_4)=(\chi-{\pi \over 2} , \chi+{\pi \over 2}, 0 , \pi ) \ , \label{eq: otoc configuration}
\end{align}
where $\chi \in (-\pi/2,\pi/2)$. Therefore, we will evaluate the following Euclidean four point function.
\begin{align}
	\cF(\chi)\equiv {\langle \Phi^\dr (\chi-\pi/2,\chi+\pi/2)\Phi^\dr (0,\pi) \rangle\over \Phi_{cl}(\chi-\pi/2,\chi+\pi/2)\Phi_{cl}(0,\pi)}=\cF_d+g^2 \mathcal{F}^{(1)}(\chi)+ g^4\mathcal{F}^{(2)}(\chi)+\mathcal{O}(g^6) \ ,
\end{align}
where the numerical value $\cF_d=[\Phi^\dr_0(-{\pi \over 2})]^2$ corresponds to the disconnected diagrams.

\begin{figure}[t!]
\centering
\begin{subfloat}[][Leading $\sim \mathcal{O}(g^2)$]{\includegraphics[width=.33\linewidth]{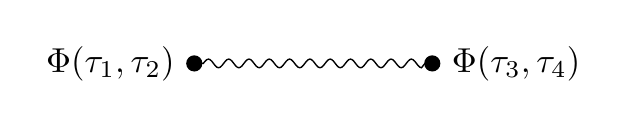}\label{fig: leading diagram}}
\end{subfloat}
\begin{subfloat}[][Loops $\sim \mathcal{O}(g^4)$]{
\includegraphics[width=.33\linewidth]{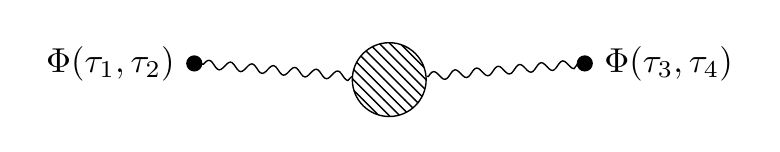}\label{fig: loops diagrams}}
\end{subfloat}\\
\begin{subfloat}[][Three soft modes scattering $\sim \mathcal{O}(g^4)$]{
\includegraphics[width=.32\linewidth]{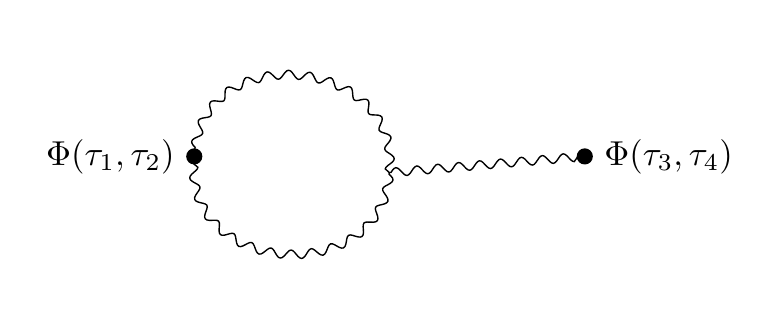}
\includegraphics[width=.32\linewidth]{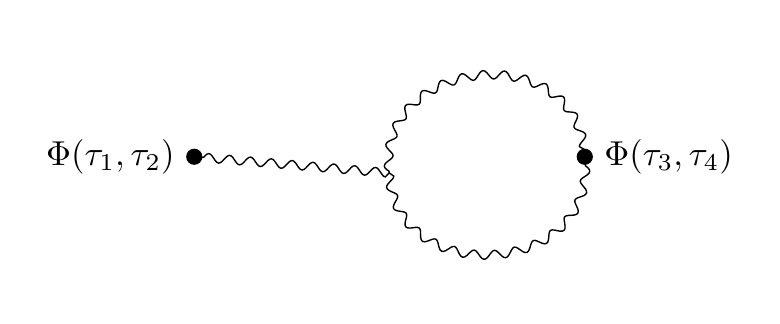}}
\end{subfloat}
\begin{subfloat}[][Two soft modes exchange $\sim~\mathcal{O}(g^4)$]{
\includegraphics[width=.31\linewidth]{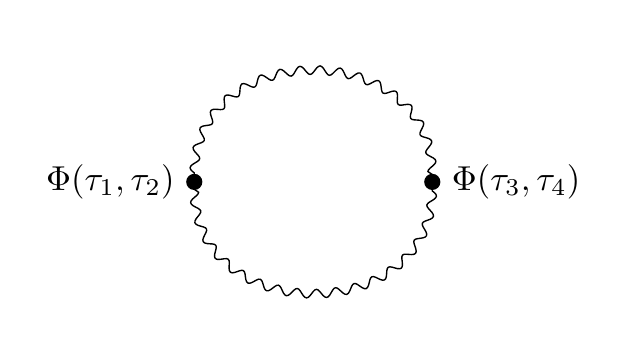}\label{fig: two soft mode exchange}}
\end{subfloat}\\
\begin{subfloat}[][Soft modes dressing $\sim \mathcal{O}(g^4)$]{
\includegraphics[width=.33\linewidth]{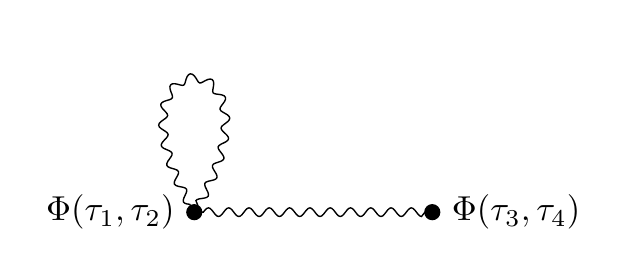}
\includegraphics[width=.33\linewidth]{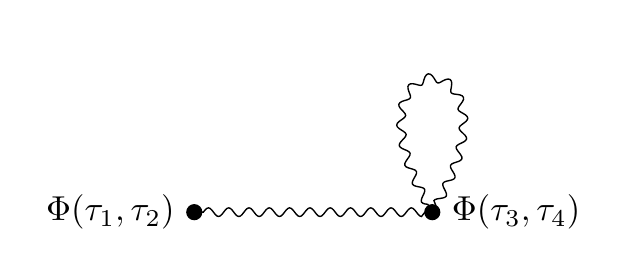}\label{fig: soft mode dressing}}
\end{subfloat}
\caption{Diagrams for the contributions of the soft mode to the Euclidean two point function of bi-local fields. We represent the dressed bi-local field by a dot.}
\label{fig: diagrams}
\end{figure}

\newpage

\noindent
{\bf Leading Connected Diagram of order $\mathcal{O}(g^2)$ :} From the soft mode expansion~\eqref{eq: soft mode expansion}, the leading contribution of the soft modes to the (connected) four point function at the configuration~\eqref{eq: otoc configuration} (See Figure~\ref{fig: leading diagram}) is found to be
\begin{equation}
	\cF^{(1)}(\chi)=\sum_{|n|\geqq 2} \langle\epsilon_{-n} \epsilon_n \rangle e^{in\chi } e^{-{n\pi i \over 2}}f^{(1)}_{-n}(-{\pi\over 2} )f^{(1)}_{n}(-{\pi\over2} )=4 h^2 \sum_{\substack{|n|\geqq 2\\ n\;:\;\text{even} }} {e^{in\chi } e^{-{\pi i n\over 2}}\over 2\pi (n^2-1) } \ ,\label{eq: leading connected diagram contribution}
\end{equation}
where we used the leading propagator of the soft mode in \eqref{eq: leading soft mode propagator}. This can be written as the following contour integral
\begin{align}
	\cF^{(1)}(\chi)=&4 h^2 {1\over 2\pi i}\oint_{\mathcal{C}} d\zeta\; { {\pi \over 2}\over \sin {\pi \zeta\over 2} } {e^{i\zeta\chi } \over 2\pi (\zeta^2-1)}\ .
\end{align}
\begin{figure}[t!]
\centering
\begin{minipage}[c]{0.4\linewidth}
\centering
\includegraphics[width=\linewidth]{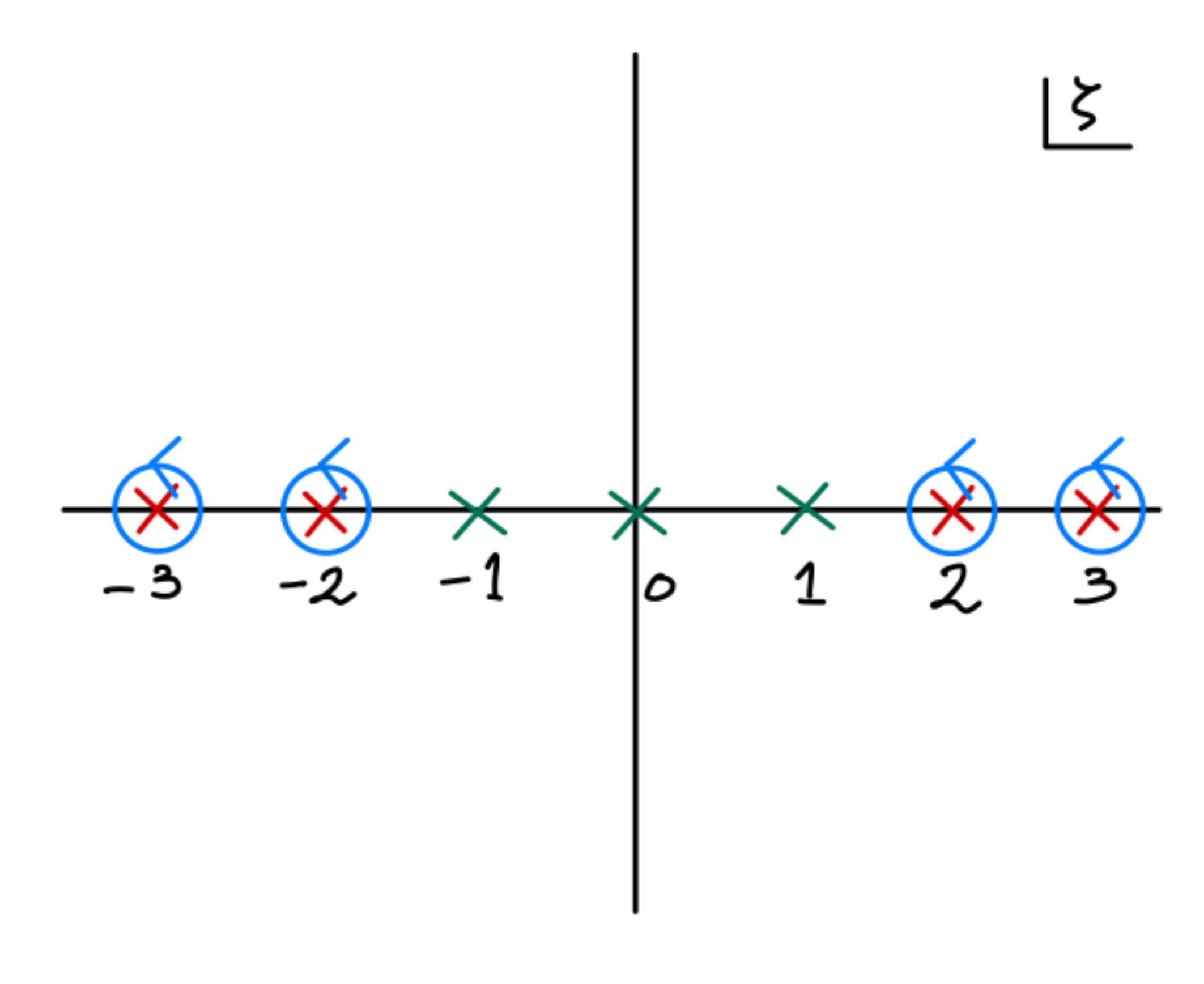}
\end{minipage}
$\quad\xRightarrow{\quad\text{Deform}\quad}\quad$
\begin{minipage}[c]{0.4\linewidth}
\centering
\includegraphics[width=\linewidth]{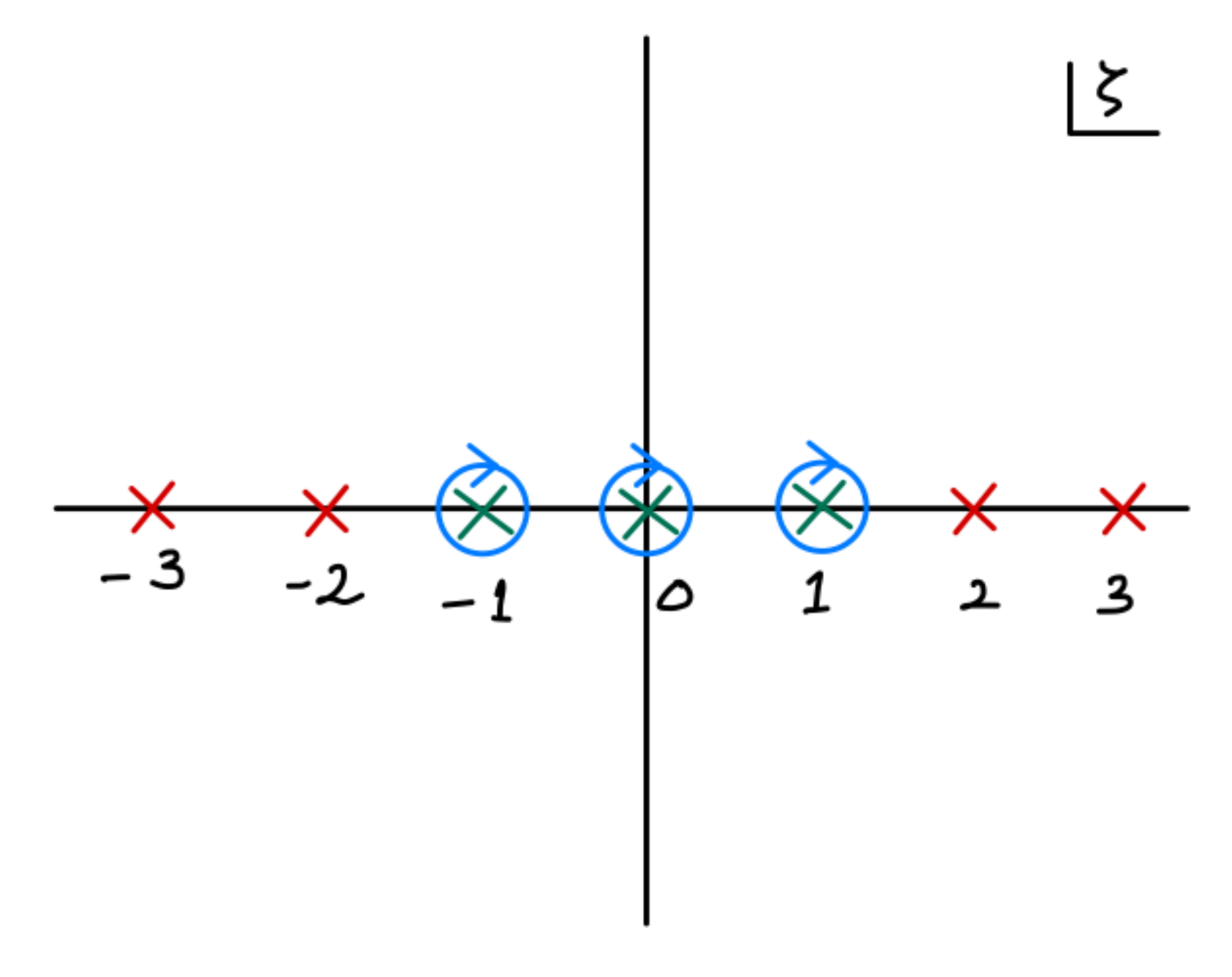}
\end{minipage}
	\caption{The contribution of the soft mode can be written as the contour integral along circles with small radius centered at $\zeta\in \mathbb{Z}/\{-1,0,1\}$. After deforming the contour, one can rewrite it as the sum of residues at $\zeta=-1,0,1$.}
	\label{fig: contour1}
\end{figure}
Here, the contour $\mathcal{C}$ is a collection of small counterclockwise circles centered at $\zeta\in\mathbb{Z}/\{ -1,0,1\}$ (See Figure~\ref{fig: contour1}). Then, we deform the contour to pick up the rest of poles. \ie 
\begin{align}
	\cF^{(1)}(\chi)=-4 h^2 \sum_{n=0,\pm 1 }\res_{\zeta=n} \; { {\pi \over 2}\over \sin {\pi \zeta\over 2} } { e^{i\zeta\chi }\over 2\pi (\zeta^2-1)}={2h^2\over \pi } \left(1- {\pi \over 2}\cos \chi\right)\ .
\end{align}

\noindent
{\bf Loop Correction  of order $\mathcal{O}(g^4)$: } Now, let us consider the loop corrections to the soft mode propagator in \eqref{eq: quartic loop correction} and \eqref{eq: cubic loop correction} (See Figure~\ref{fig: loops diagrams}), which give a contribution to $g^4\cF^{(2)}(\chi)$.
\begin{align}
	\cF^{(2)}_{\text{\tiny loop}}(\chi)=&4 h^2 \sum_{\substack{|n|\geqq 2\\ n\;:\;\text{even} }} e^{in\chi } e^{-{\pi i n\over 2}} \langle \epsilon_{-n}\epsilon_n\rangle_{\text{\tiny loop}} n^2\;\;\subset \cF^{(2)}(\chi)\ .
\end{align}
For the loop with one quartic vertex, one can repeat the same calculation as the leading contribution. However, for the loop with two cubic vertices, one has to calculate $|n|> 2$ case and $n=\pm 2$ case separately. For $|n|> 2$, one can also rewrite the contribution as a contour integral where the contour is a collection of small counterclockwise circles centered at $\zeta\in \mathbb{Z}/\{-2,-1,0,1,2\}$. And, the deformation of the contour gives the sum of residues at $\zeta=-2,-1,0,1,2$. Summing up the results for $|n|>2$ and $n=\pm 2$, we have
\begin{align}
	\cF^{(2)}_{\text{\tiny loop}}(\chi)=&{7  h^2\over 2 \pi^2}\left[ -1+{\pi\over 168}(88-3\pi^2 +12\chi^2)\cos \chi  + {1\over 126} (-79 +9\pi^2 +36 \chi^2)\cos 2\chi\right. \cr
	&\hspace{20mm}\left. +{5\pi  \over 14} \chi \sin \chi - {2\over 3}\chi \sin 2\chi\right]\ .\label{eq: euclidean loop contribution}
\end{align}

\noindent
{\bf Correction from Scattering of Three Soft Modes with Cubic Vertex of order $\mathcal{O}(g^4)$:} $\cF^{(2)}(\chi)$ includes the contribution from the scattering of three soft modes emitted from the bi-locals. Using the cubic vertex in ~\eqref{eq: cubi action momentum} and the soft mode eigenfunction $f^{(1)}_n(\sigma)$ in \eqref{eq: soft mode1 at special sigma} and $f^{(2)}_{m,n}(\sigma)$ in \eqref{eq: soft mode2 at special sigma}, this can be written as follows.
\begin{align}
	&\cF^{(2)}_{\text{\tiny three soft modes scattering}}\cr
	=&- {2h^2\over \pi^2} \sum_{\substack{m,n\;:\;\text{\tiny even} \\ |m|,|n|\geqq 2, |m+n|\geqq 2}} {(2h-1)(m^2+mn+n^2)\over (n^2-1)(m^2-1)[(n+m)^2-1]} (-1)^{n+m\over 2}\cos(n+m)\chi\cr
	&- {2h^2\over \pi^2}\sum_{\substack{m,n\;:\;\text{\tiny odd} \\ |m|,|n|\geqq 2, |m+n|\geqq 2}} {(mn+1)(m^2+mn+n^2) \over nm (n^2-1)(m^2-1)[(n+m)^2-1]} (-1)^{n+m\over 2}\cos(n+m)\chi\ .
\end{align}
\begin{figure}[t!]
\centering
\begin{subfloat}[][$n\ne\pm 2$]{
\begin{minipage}[c]{0.4\linewidth}
\centering
\includegraphics[width=\linewidth]{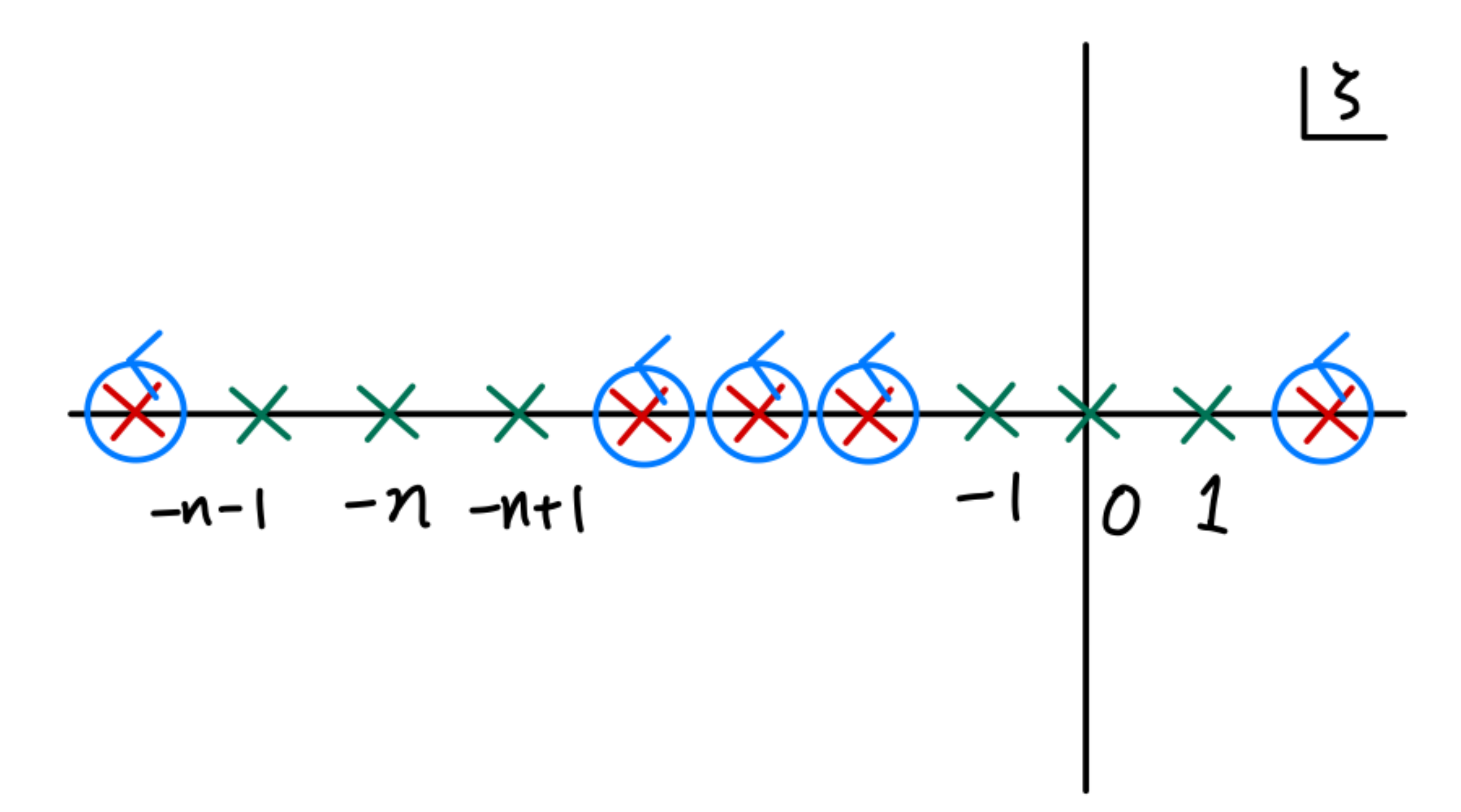}
\end{minipage}
$\quad\xRightarrow{\quad\text{deform}\quad}\quad$
\begin{minipage}[c]{0.4\linewidth}
\centering
\includegraphics[width=\linewidth]{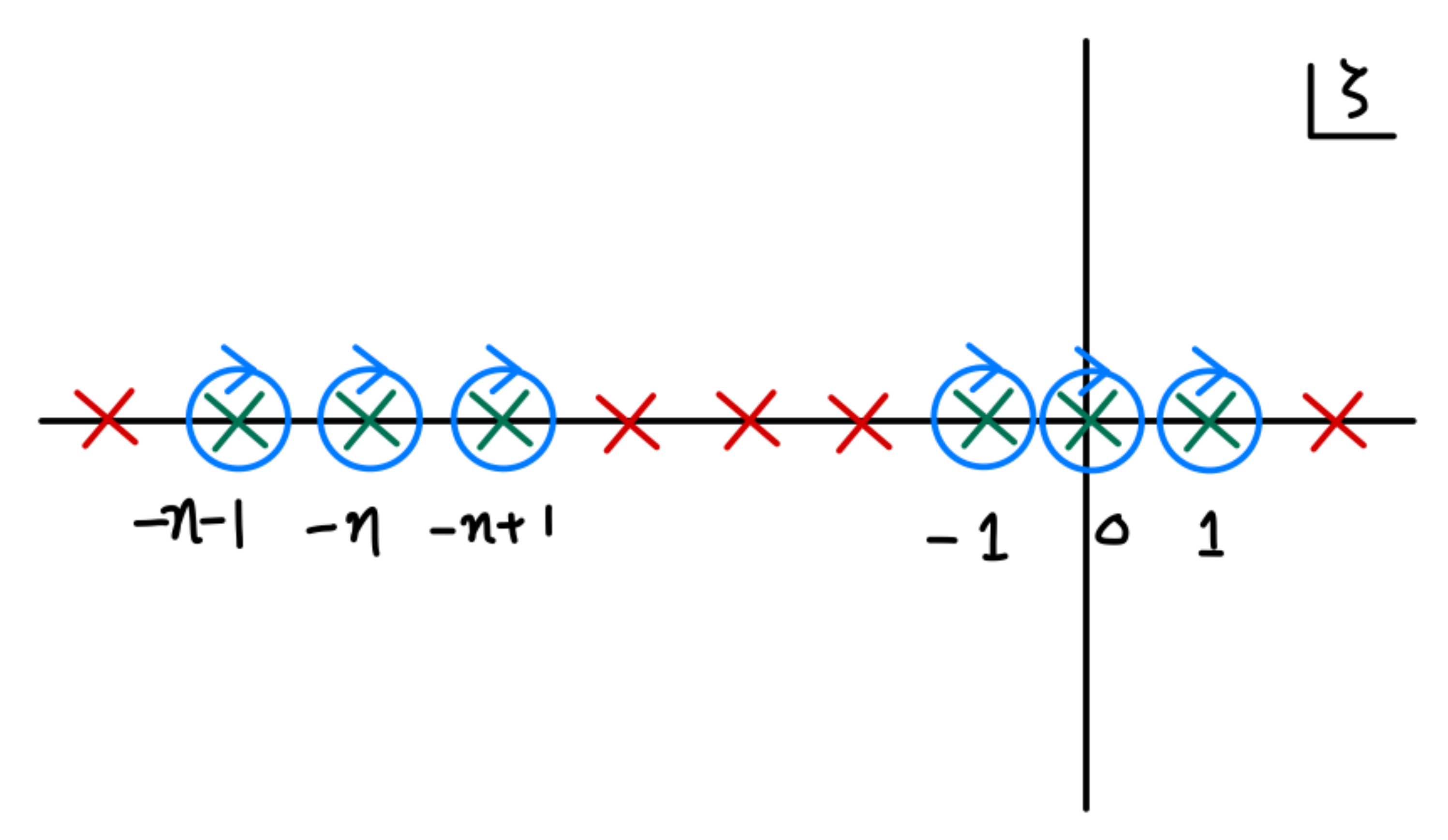}
\end{minipage}}
\end{subfloat}\\
\begin{subfloat}[][$n= 2$]{
\begin{minipage}[c]{0.4\linewidth}
\centering
\includegraphics[width=\linewidth]{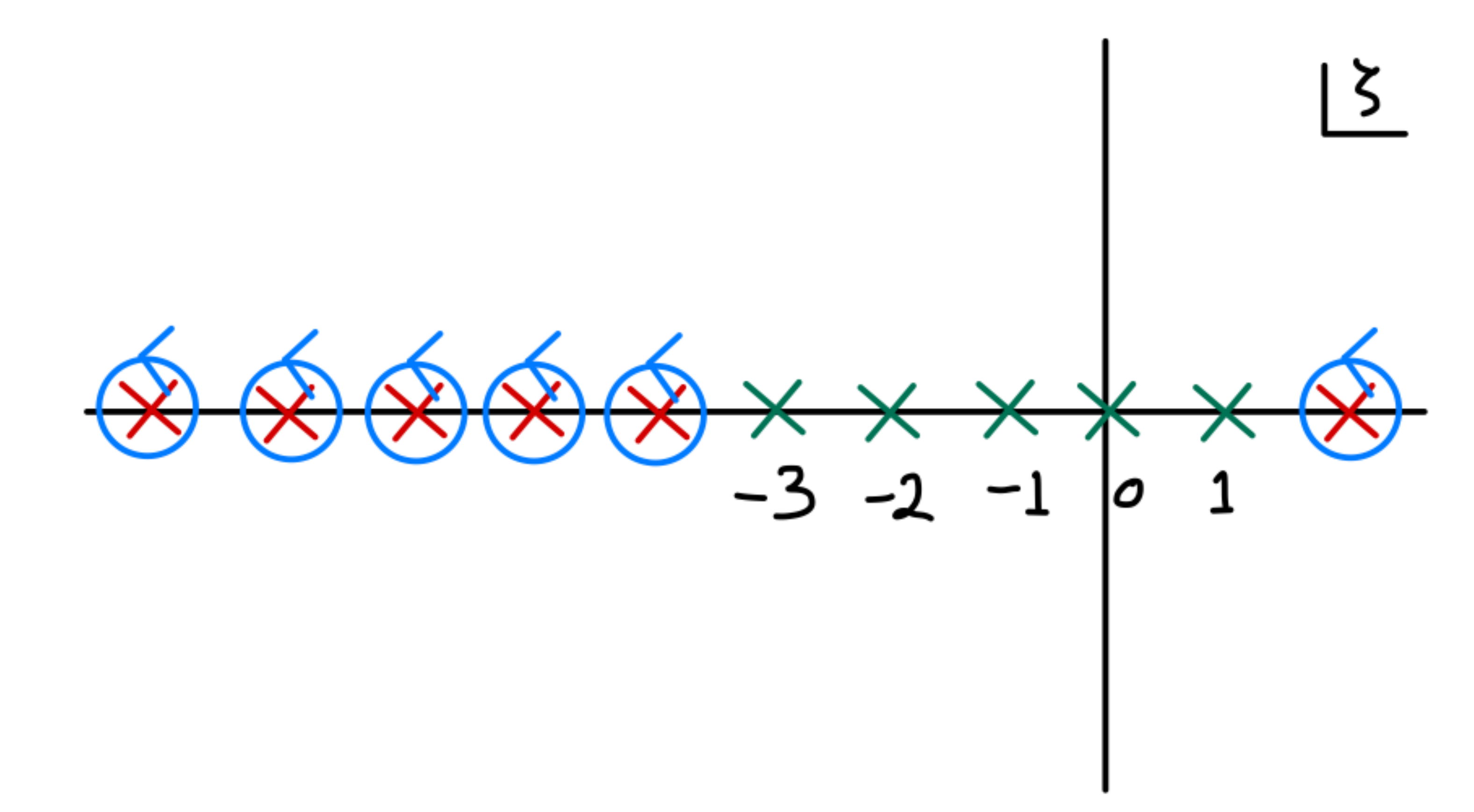}
\end{minipage}
$\quad\xRightarrow{\quad\text{deform}\quad}\quad$
\begin{minipage}[c]{0.4\linewidth}
\centering
\includegraphics[width=\linewidth]{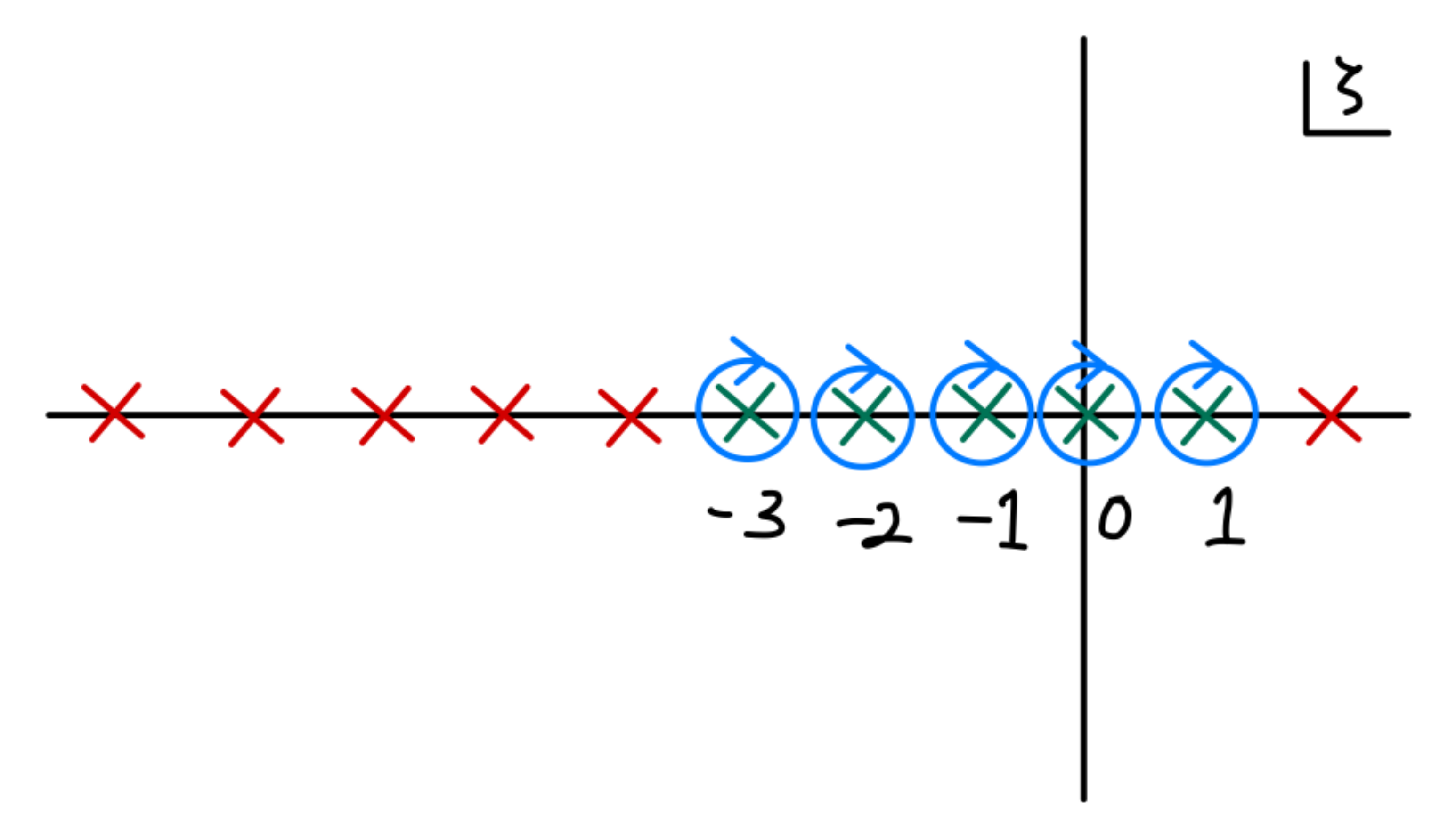}
\end{minipage}}
\end{subfloat}
	\caption{The soft mode contribution can be expressed as the contour integral along circles with small radius centered at $\zeta\in\mathbb{Z}/\{-1,0,1,-n-1,-n,-n+1\}$ for $n\ne \pm 2$, at $\zeta\in\mathbb{Z}/\{-3,-2-1,0,1,\}$ for $n=2$ or at $\zeta\in\mathbb{Z}/\{-1,0,1,2,3\}$ for $n=-2$, respectively. By deforming the contour, it can be expressed as the sum of residues at $\zeta=-n-1,-n,-n+1,-1,0,1$, $\zeta=-3,-2,-1,0,1$ or $\zeta=-1,0,1,2,3$, respectively.}
	\label{fig: contour2}
\end{figure}
As in the calculation of the loop with two cubic vertices, one has to evaluate them for $|n|> 2$ and $n=\pm 2$ separately because of the constraint $|m+n|\geqq 2$ in the summation (See Figure~\ref{fig: contour2}). We found
\begin{align}
	&\cF^{(2)}_{\text{\tiny three soft modes scattering}}={h^2\over \pi^2}\left[-1 +{1\over 36}(178+9(2h-1)\pi^2-72\chi^2)\cos 2\chi \right.\cr
	&\hspace{25mm}\left.+{\pi \over 8} (-16h +\pi^2 -4\chi^2)\cos \chi +\pi (2h-1) \chi \sin \chi +{16\over 3}\chi \sin  2\chi \right]\ .\label{eq: euclidean three soft modes}
\end{align}

\noindent
{\bf Correction from Two Soft Modes Exchanges of order $\mathcal{O}(g^4)$:} The exchange of two soft modes gives the contribution of order $\mathcal{O}(g^4)$ (See Figure~\ref{fig: two soft mode exchange}). For this, using the soft mode eigenfunction $f^{(2)}_{m,n} (\sigma)$ in \eqref{eq: soft mode2 at special sigma}, one can evaluate the diagram for the two soft mode exchange in a similar way.
\begin{align}
	&\cF^{(2)}_{\text{\tiny two soft modes exchange}}\cr
	=&{h^2(2h-1)^2\over 2\pi^2}\sum_{\substack{n,m\;:\;\text{even}\\ |m|,|n|\geqq 2}} {e^{-i(m+n)\chi}e^{{\pi i \over 2}(m+n)}\over (n^2-1)(m^2-1)} +{h^2\over 2\pi^2}\sum_{\substack{n,m\;:\;\text{odd}\\ |m|,|n|\geqq 2}} {(mn+1)^2 e^{-i(m+n)\chi}e^{{\pi i \over 2}(m+n)}\over n^2m^2(n^2-1)(m^2-1)}\cr
	=&{h^2\over 8\pi^2} \left[4 +\pi^2 (-2+\chi^2) +4\pi(3+\chi^2)\cos \chi + (2h-1)^2(-2+\pi\cos\chi)^2\right.\cr
	&\hspace{10mm} \left. +(-22+8\chi^2)\cos 2\chi - 2\pi \chi \sin \chi -24\chi \sin 2\chi  \right]\ . \label{eq: euclidean two soft modes}
\end{align}

\noindent
{\bf Correction from Soft Mode Dressing of order $\mathcal{O}(g^4)$:} One can also consider a correction to the leading diagram in Figure~\ref{fig: leading diagram} by the soft mode dressing. (See Figure~\ref{fig: soft mode dressing}). This contribution of order $\mathcal{O}(g^4)$ can be written as
\begin{align}
	\cF^{(2)}_{\text{\tiny soft mode dressing}}=2\sum_{|n|,|m|\geqq 2} e^{in\chi}e^{-{n\pi i \over 2}} f^{(3)}_{m,n,-m}(-{\pi \over 2}) f^{(1)}_{-n}(-{\pi \over 2})\langle\epsilon_{-n}\epsilon_n \rangle\langle\epsilon_{-m}\epsilon_m \rangle\ .
\end{align}
In the same way as before, the contour integral representation for the summation over $m$ gives the leading connected diagram in \eqref{eq: leading connected diagram contribution} (but, of order $\mathcal{O}(g^4)$) up to numerical constant. \ie
\begin{align}
	\cF^{(2)}_{\text{\tiny soft mode dressing}}=& {h\over 36\pi}(3+6h-\pi^2) \sum_{|n|\geqq 2} e^{i n\chi } f^{(1)}_{-n}(-{\pi \over 2})f^{(1)}_{n}(-{\pi \over 2})\langle\epsilon_{-n}\epsilon_n \rangle\ .
\end{align}
Hence, we have
\begin{align}
	\cF^{(2)}_{\text{\tiny soft mode dressing}}=& {h^3\over 18\pi^2}(3+6h-\pi^2)\left(1-{\pi \over 2}\cos \chi \right)\ . 
\end{align}

\noindent
{\bf Total Contribution of Soft Modes:} Summing up the above all results, we have
\begin{align}
	&\cF_{\text{\tiny total}}(\chi)\cr
	=&\cF_d+g^2\cF^{(1)}+g^4\left[\cF^{(2)}_{\text{\tiny loop}}+\cF^{(2)}_{\text{\tiny three soft modes scattering}}+\cF^{(2)}_{\text{\tiny two soft modes exchange}}+\cF^{(2)}_{\text{\tiny soft mode dressing}}\right]\cr
	=&\cF_d+ {2g^2h^2\over \pi } \left( 1- {\pi \over 2}\cos \chi\right)\cr
	&+{g^4 h^2\over 144\pi^2}\bigg[288 \pi h \chi \sin \chi + 9\pi^2 (1+2h)^2\cos 2\chi  + \pi \big[36\chi^2 - 4h(3+78h -\pi^2) +9(24+\pi^2)\big]\cos \chi \cr
	&\hspace{20mm}  +18 \pi^2 \chi^2+12 h^2(28+3\pi^2)-44h(6+\pi^2)-504-27\pi^2  \bigg]\ .\label{eq: euclidean four point function total}
\end{align}
%

\subsection{Real-time Out-of-time-ordered Correlator}
\label{sec: real time otoc}

\begin{figure}[t!]
\centering
\begin{minipage}[c]{0.3\linewidth}
\centering
\includegraphics[width=\linewidth]{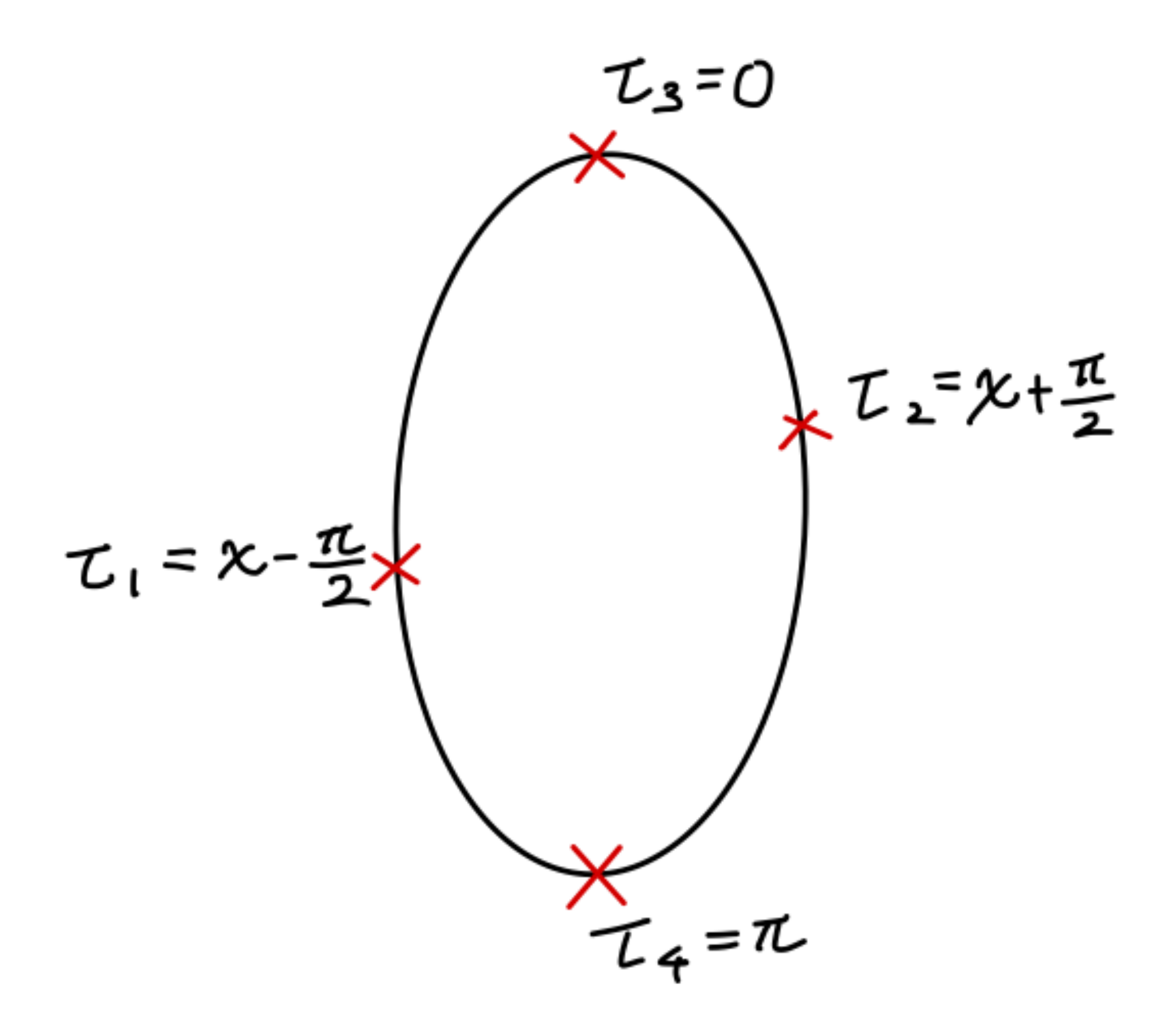}
\end{minipage}
$\quad\xRightarrow{\quad\text{Analytic continuation}\quad}\quad$
\begin{minipage}[c]{0.3\linewidth}
\centering
\includegraphics[width=\linewidth]{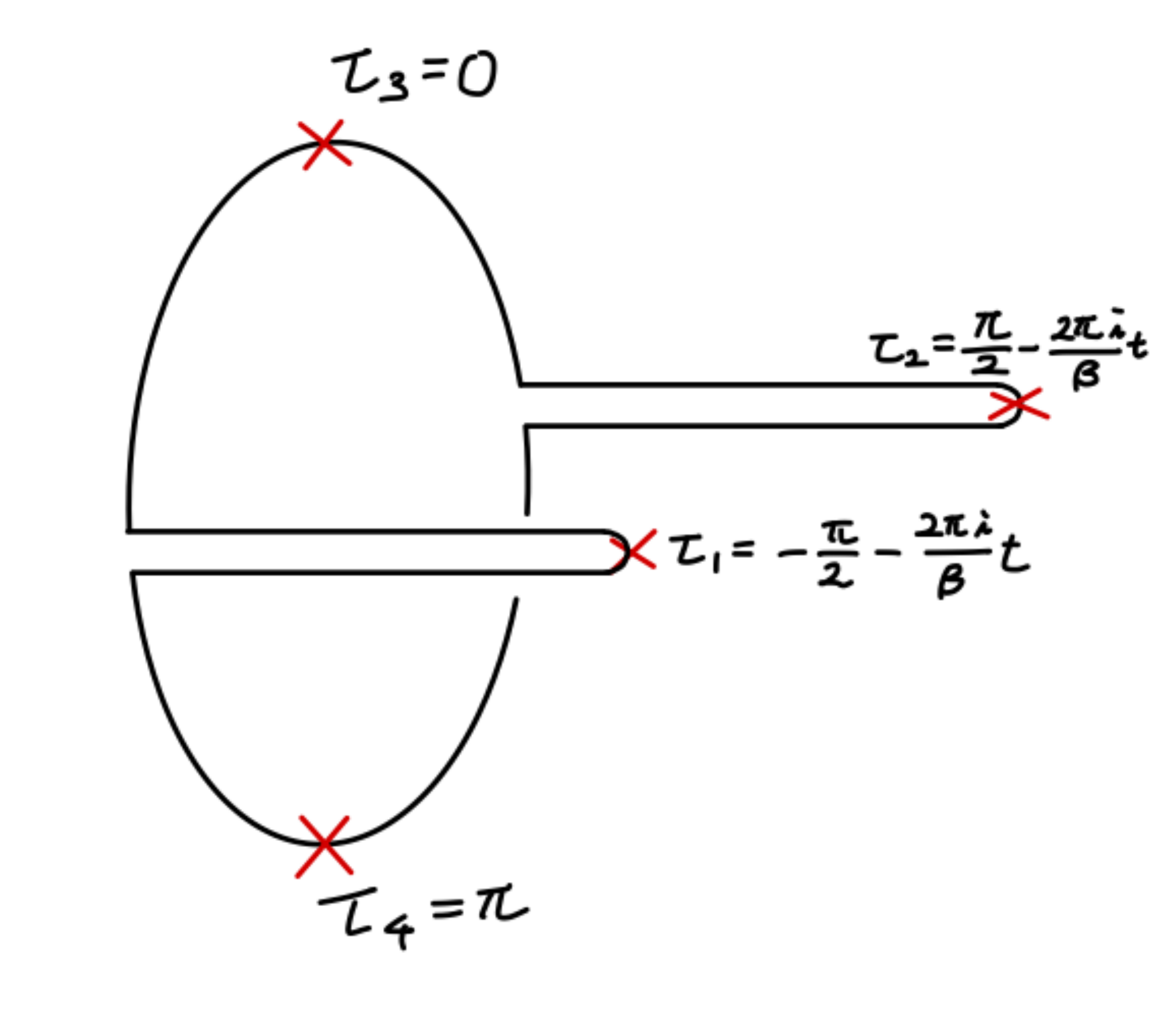}
\end{minipage}
	\caption{Analytic continuation from the configuration on Euclidean thermal circle to the real time out-of-time-ordered configuration.}
	\label{fig: analytic continuation}
\end{figure}
From the Euclidean four point function in~\eqref{eq: euclidean four point function total}, we take the analytic continuation from Euclidean time $\chi$ to Lorentzian time $t$ (See Figure~\ref{fig: analytic continuation})
\begin{equation}
	\chi \quad\longrightarrow \quad -{2\pi i t\over \beta}\ ,
\end{equation}
in order to obtain the OTOC:
\begin{align}
	\cF_{\text{\tiny OTOC}}(t)=&\cF_d- g^2 \left[{h^2\over 2}e^{2\pi t \over \beta}- {2g^2 h^2\over \pi} +{h^2\over 2}e^{-{2\pi t \over \beta}} \right] \cr
	&\hspace{5mm} +g^4 \bigg[  {h^2(2h+1)^2\over 32}e^{4\pi t\over \beta} -{h^2\pi \over 2 \beta^2 }  t^2 e^{2\pi t\over \beta} - {2 h^3\over \beta} t e^{2\pi t\over \beta} \cr
	&\hspace{20mm}+ {h^2(312h^2 -4h(\pi^2-3)-9(\pi^2+24))\over 288\pi }e^{2\pi t\over \beta}+\cdots \bigg]\ ,\label{eq: otoc total}
\end{align}
where the ellipsis represents terms that do not grow exponentially at order $\mathcal{O}(g^4)$.

First of all, note that the leading Lyapunov exponent saturates the bound on chaos~\cite{Maldacena:2016hyu,Maldacena:2016upp,Yoon:2019cql}. \ie
\begin{equation}
	\lambda_L={2\pi \over \beta} +\mathcal{O}(g^2)\ .
\end{equation}
Among the contribution of order $\mathcal{O}(g^4)$ in \eqref{eq: otoc total}, a term like $-{2h^3\over \beta }g^4 t e^{2\pi t \over \beta}$ has been observed to give a correction to the maximum Lyapunov exponent. In SYK model, for instance, the contribution of the non-zero mode to OTOC gives $t e^{2\pi t \over \beta}$ which leads to the $1/\beta J$ correction to the leading Lyapunov exponent~\cite{Maldacena:2016hyu}. While the $1/\beta J$ correction in SYK model decreases the Lyapunov exponent from the maximum value ${2\pi \over \beta}$, the $1/c$ correction to the Lyapunov exponent from the Virasoro conformal block in large $c$ was shown~\cite{Fitzpatrick:2016thx,Chen:2016cms} to increase the Lyapunov exponent. \ie $\lambda_L={2\pi \over \beta}(1+{12\over c})$. In our result, the contribution $-{2h^3\over \beta }g^4 t e^{2\pi t \over \beta}$ in~\eqref{eq: otoc total} seemingly plays a role of increasing the Lyapunov exponent. However,  we have other terms at order $\mathcal{O}(g^4)$ which grow faster than $t e^{2\pi t \over \beta}$. In particular, the fastest growing term, $ {h^2(2h+1)^2\over 32}e^{4\pi t\over \beta} $ in \eqref{eq: otoc total}, naively seems to violate the bound on chaos because it grows exponentially with growth rate ${4\pi \over \beta}$. However, it turns out that it reduces the Lyapunov exponent because its contribution to the OTOC has opposite sign to the leading exponential growth. 

Note that each contribution of order $\mathcal{O}(g^4)$ (\eg the analytic continuation of \eqref{eq: euclidean loop contribution}, \eqref{eq: euclidean three soft modes}, \eqref{eq: euclidean two soft modes}) includes exponentially growing terms such as $t^2 e^{4\pi t\over \beta}$ and $t e^{4\pi t\over \beta}$ which grow faster than $e^{4\pi t\over \beta}$. In particular, those in \eqref{eq: euclidean loop contribution} and \eqref{eq: euclidean two soft modes} play a role of increasing the Lyapunov exponent. On the other hand, the analogous terms in \eqref{eq: euclidean three soft modes} decreases the Lyapunov exponent. It is interesting that those fast growing contributions are cancelled exactly, and in the end the remaining fastest growth $e^{4\pi t\over \beta}$ decreases the Lyapunov exponent as we have seen. If we considered only the loop correction in \eqref{eq: euclidean loop contribution} for the calculation of OTOC, we would get
\begin{align}
	\cF_{\text{\tiny OTOC, loop}}(t)=&\cF_d- g^2 \left[{h^2\over 2}e^{2\pi t \over \beta}- {2g^2 h^2\over \pi} +{h^2\over 2}e^{-{2\pi t \over \beta}} \right]  \cr
	&\hspace{5mm}+g^4\bigg[-{2h^2 \over \beta^2 }t^2e^{4\pi t\over \beta} +{7h^2 \over 3\pi \beta}te^{4\pi t\over \beta}+{h^2 (-79+9\pi^2)\over 72\pi^2}e^{4\pi t\over \beta} -{h^2 \pi \over 2\beta^2}t^2 e^{2\pi t\over \beta}  \cr
	&\hspace{20mm} -{5h^2 \over 4\beta}t e^{2\pi t\over \beta} -{h^2(-88+3\pi^2)\over 96\pi }e^{2\pi t\over \beta} +\cdots \bigg]\ .\label{eq: otoc loop}
\end{align}
One can easily see that the fastest exponential growth at of order $\mathcal{O}(g^4)$ increases the Lyapunov exponent.

\begin{figure}[t!]
\centering
\begin{subfloat}[][Total Contribution]{
\includegraphics[width=.48\linewidth]{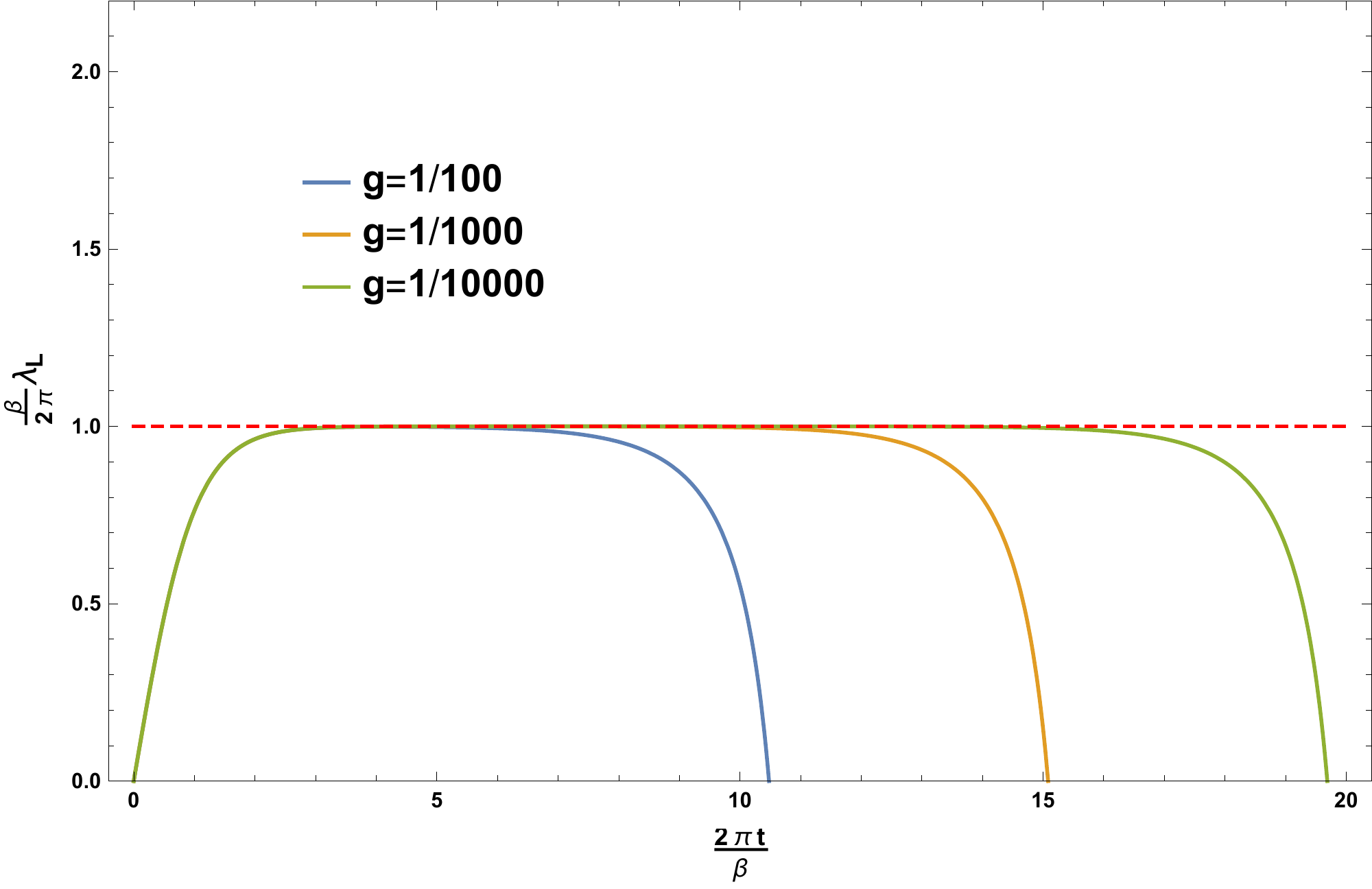}\label{fig: lyapunov exponent total contribution}}
\end{subfloat}
\begin{subfloat}[][Loop Contribution]{
\includegraphics[width=.48\linewidth]{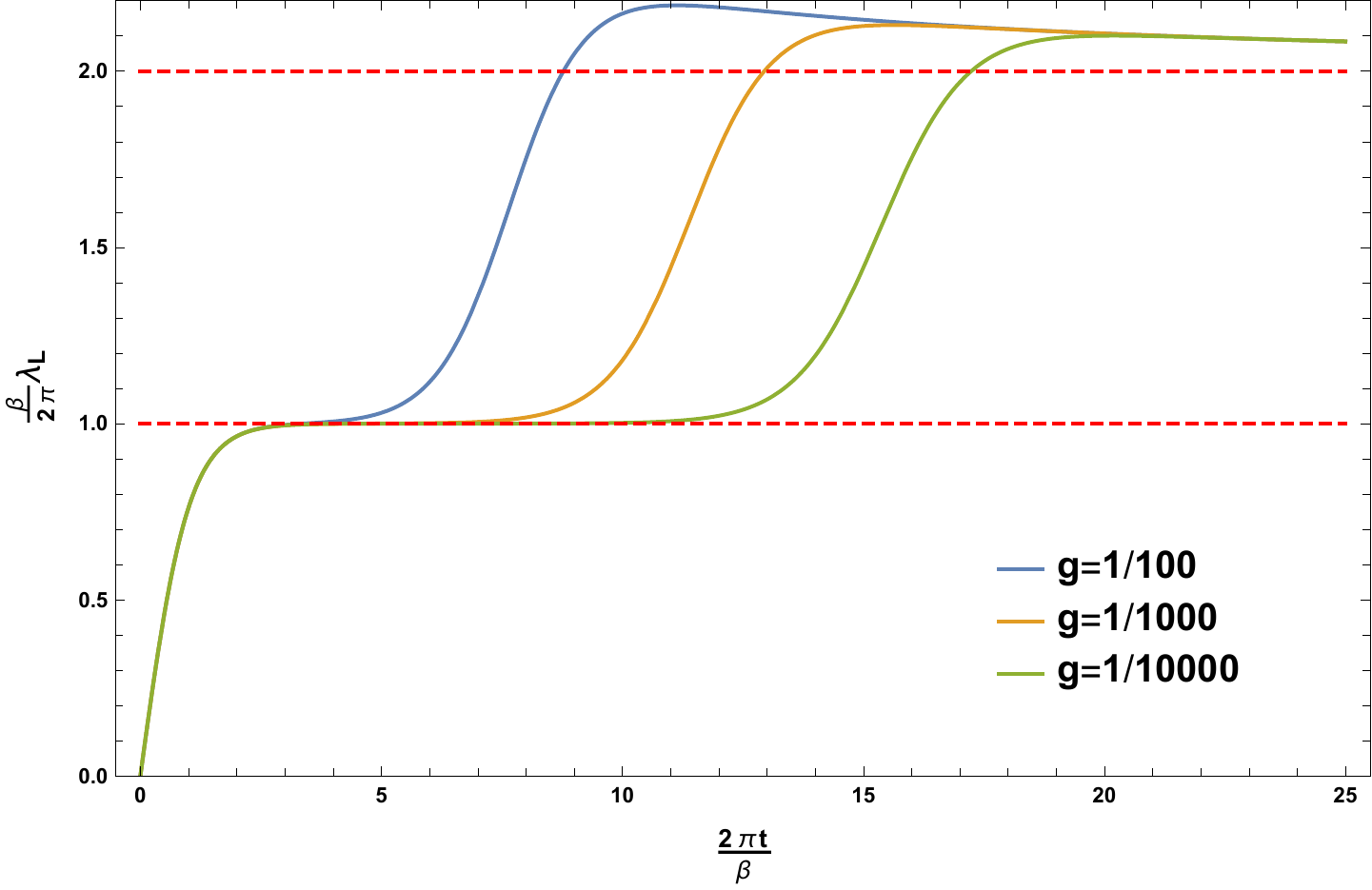}\label{fig: lyapunov exponent loop correction}}
\end{subfloat}

	\caption{Lyapunov exponent $\lambda_L(t)$ from the total contribution to OTOC and from the loop contribution to OTOC. We plot them for the case of $g=1/100, 1/1000, 1/10000$, and the corresponding scrambling time would be ${2\pi \over \beta} t\sim \log g^2 \simeq 9.21, 13.82, 18.42$. }
	\label{fig: numerical plot for lyapunov exponent}
\end{figure} 
To see the change of the Lyapunov exponent concretely, we go back to the original statement of the bound on chaos where we define the Lyapunov exponent by
\begin{equation}
	\lambda_L (t)\; \equiv \; {d\over dt}\bigg[ \log \Big(\cF_{\text{\tiny const}}-\cF(t)\Big)\bigg]   \ , 
\end{equation}
where $\cF_{\text{\tiny const}}$ corresponds to the constant terms\footnote{In the bound on chaos of~\cite{Maldacena:2015waa}, $\cF_d$ was used for this constant. However, we choose $\cF_{\text{\tiny const}}$ to be all constant terms in all order $g$. For example, $\cF_{\text{\tiny const}}=\cF_d+{2g^2 h^2\over \pi} +\mathcal{O}(g^4) $.} in $\cF(t)$. The bound on chaos states
\begin{equation}
	\lambda_L (t) \; \leqq\;  {2\pi \over \beta}\hspace{10mm}\mbox{for }\quad \beta< t< t_\ast\ ,\label{eq: def of lyapunov exponent}
\end{equation}
where $t_\ast\sim {\beta\over 2\pi}\log {1\over g^2}$ is the scrambling time.

We plot numerically the Lyapunov exponent $\lambda_L(t)$ as a function of time from $\cF_{\text{\tiny OTOC}}(t)$ in \eqref{eq: otoc total} and from $\cF_{\text{\tiny OTOC, loop}}(t)$ in \eqref{eq: otoc loop}, respectively. See Figure~\ref{fig: numerical plot for lyapunov exponent}. Here, we plot the Lyapunov exponent $\lambda_L(t)$ for $g=1/100, 1/1000, 1/10000$ of which the scrambling time would be ${2\pi \over \beta} t\sim \log g^2 \simeq 9.21, 13.82, 18.42$, respectively. In Figure~\ref{fig: lyapunov exponent total contribution}, the Lyapunov exponent $\lambda_L(t)$ from $\cF_{\text{\tiny OTOC}}(t)$ in \eqref{eq: otoc total} is less than ${2\pi \over  \beta}$. As time increase, the Lyapunov exponent quickly saturates the bound, and it begins to decrease around the scrambling time. The term ${g^4h^2(2h+1)^2\over 32}e^{4\pi t\over \beta}$ at order $\mathcal{O}(g^4)$ in \eqref{eq: otoc total} is responsible for this decrease of Lyapunov exponent. The Lyapunov exponent seemingly vanishes beyond the scrambling time $t_\ast$. However, we cannot trust the Lyapunov exponent beyond the scrambling time because the small $g$ perturbation will break down.

On the other hand, if we had only the loop correction for the quantum correction to the OTOC, we would observe the violation of chaos bound before the scrambling time $t_\ast$. See Figure~\ref{fig: lyapunov exponent loop correction}. This violation of the bound mainly comes from the fastest exponential growth $-g^4{2h^2 \over \beta^2 }t^2e^{4\pi t\over \beta} $ at order $\mathcal{O}(g^4)$ in \eqref{eq: otoc loop}.

\section{Conclusion}
\label{sec:conclusion}

In this paper, we have evaluated the quantum correction of order $\mathcal{O}(g^4)$ by the Schwarzian soft mode to the OTOC. As is well known, the OTOC at order $\mathcal{O}(g^2)$ grows exponentially with the maximum growth rate ${2\pi \over \beta}$~\cite{Maldacena:2016hyu,Maldacena:2016upp,Yoon:2019cql}. At order $\mathcal{O}(g^4)$, we have found that the loop correction by the Schwarzian soft modes and the correction by two soft mode exchanges make the OTOC grow faster than the maximal growth. On the other hand, the correction by three soft mode scattering decreases the exponential growth rate. And, we have showed that the total correction slows down the exponential growth of the OTOC.

It is important to issue caveats in our analysis. First of all, we have not shown that the chaos bound would hold beyond the scrambling time, but we have found that the soft mode contribution to the OTOC at order $\mathcal{O}(g^4)$ slows down the exponential growth of order $\mathcal{O}(g^2)$. To see the behavior of the OTOC beyond the scrambling time, one need to go beyond the perturbation, or, at least, the higher order corrections are required to estimate the behavior beyond the scrambling time. For example, $\mathcal{O}(g^6)$ might have $e^{6\pi t\over \beta}$ growth, and depending on its sign the behavior around the scrambling time might be changed. We hope to report the higher order calculations in near future. Also, it would be highly interesting to find a constraint on the behavior of OTOC beyond the scrambling time from a simple physical argument.

In addition, our analysis is based on the Schwarzian theory which might not be as universal as other approaches such as the Virasoro conformal block~\cite{Fitzpatrick:2016thx,Chen:2016cms,Fitzpatrick:2016mjq} or pole-skipping phenomenon~\cite{Grozdanov:2017ajz,Blake:2017ris,Haehl:2018izb,Blake:2018leo,Grozdanov:2018kkt,Grozdanov:2019uhi,Blake:2019otz}. Nevertheless, the low energy physics of many interesting models such as SYK-like models and the dilaton gravity on nearly AdS$_2$ is described by Schwarzian action, in which our result can provide the understanding of the quantum correction to the chaos. It is interesting to explore the quantum correction to chaos in the context of ``pole-skipping'' phenomenon in the CFT$_2$ or higher dimensional CFT.

Finally, we have not evaluated all of the OTOC at order $\mathcal{O}(g^4)$, but we have calculated the contribution of the Schwarzian soft mode at order $\mathcal{O}(g^4)$. We have not considered the interaction of the matter fields which could also give a contribution to the OTOC. Unlike the soft mode, the contribution of matter fields might not be universal, but it would possibly depend on the details of models. Nevertheless, it would be interesting if the quantum correction to the chaos can constraint the matter interaction.

\acknowledgments

We would like to thank Piljin Yi, Keun-Young Kim, Viktor Jahnke, Mitsuhiro Nishida, Martin Ammon and Robert de Mello Koch for extensive discussions. JY thanks the Erwin Schrodinger International Institute (ESI) where this work was initiated during the program ``Higher Spins and Holography 2019''. JY thank the Okinawa Institute of Science and Technology~(OIST), the Gwangju Institute of Science and Technology~(GIST) and the South China Normal University~(SCNU) for the hospitality and generous support.





\bibliographystyle{JHEP}
\bibliography{correction.bib}

\end{document}